\documentstyle[aps,epsf]{revtex}
\draft
\begin{document}
\title{Probing the Vacuum Induced Coherence in a $\Lambda$-system}
\author{Sunish Menon and G. S. Agarwal\footnote {Also at Jawaharlal Nehru 
Centre for Adavanced Scientific Research, Banglore, India.}}
\address{Physical Research Laboratory, Navrangpura, Ahmedabad 380 009, India.}
\date{\today}
\maketitle
\begin{abstract}
We propose a simple test to demonstrate and detect
the presence of vacuum induced coherence in a $\Lambda$-system.  We show that
the probe field absorption is modulated due to the presence of such a coherence
 which is unobservable in fluorescence.  We present analytical and 
 numerical results for the modulated absorption, the cosine and sine components 
 of which display different types of behavior.
\end{abstract}
%%\newpage

\section{introduction}
The vacuum of electromagnetic field is known \cite{gsa1} to give rise to several
types of very interesting coherence effects.  For example it gives rise to
atom-atom correlations and the collective effects.  In a single multilevel atom,
it also gives rise to coherences among levels.  The coherences are especially
significant if the relevant levels are near by.  In Ref. \cite{gsa1} it was
shown that the vacuum of the electromagnetic field can lead to the possibility
of a trapped state in a degenerate V-system.  The Ref. \cite{gsa1} also analyzed
the origin of such a trapped state in terms of suitably defined coupled and
uncoupled states.  Spontaneous emission produced a coherence between the two excited
states of the V-system. 
  Further, it was
shown that such a vacuum induced coherence (VIC) can suppress the
steady state resonance fluorescence  
\cite{stroud}, and can substantially modify the emission spectrum \cite{agassi}.
Recently, interest in this subject has been revived due to the various
possibilities of manipulating atomic properties using atomic coherence effects
\cite{{plenio},{zhu},{xia},{gsa2},{swain},{imamoglu},{java},{me},{nar},{paspalakis}}.
Using this kind of coherence effect,
Hegerfeldt and Plenio showed that periodic dark states and quantum 
beats appear in a near-degenerate V-system \cite{plenio}.   The work of Zhu, Scully
and coworkers \cite{zhu}
demonstrate that even
spectral line elimination and spontaneous emission cancellation is possible.
This was observed experimentally by Xia, Ye and Zhu in sodium dimers
\cite{xia}.  An appealing physical
picture to explain spontaneous emission cancellation was provided by 
Agarwal \cite{gsa2}.  The effect of vacuum induced coherence (VIC) on
spontaneous emission has been suggested to achieve gain without 
inversion and sub-natural line-widths \cite{swain}.  We mention that such a 
coherence mechanism is also known to occur in quantum well structures \cite{imamoglu}.
Recent studies have also shown that
vacuum induced coherence effects give rise to phase sensitive 
absorption \cite{me} and emission \cite{nar} profiles as well as 
to obtain phase control of spontaneous emission in V-systems \cite{paspalakis}.

While much of the work has been in connection with V-systems, other level schemes
like $\Lambda$-systems \cite{{java},{me},{nar}} and $\Xi$-systems \cite{scully}
have also been studied.  In case of $\Lambda$-systems, 
the coherence is produced in the ground state.  We study in this article the origin
of this coherence and the question of a proper probe for such a coherence.

The organization of this paper is as follow.  In Sec. II we show the reason
behind the origin of VIC.  In Sec. III we derive the spontaneous emission
spectrum in the presence of VIC, and show that the spectrum is independent
of VIC.  In Sec. IV we show that absorption of a weak field, as a probe
for VIC, will be uniquely modulated due to VIC.  Using second order
perturbation theory we derive the analytical results which well explains
our numerical results.  Finally, in Sec. V we present
concluding remarks.

\section{origin of vic in $\Lambda$-systems}

\begin{figure}[h]
\hspace*{3.5 cm}
%\vspace*{}
%\epsfxsize  4.5 in
%\epsfysize 2.5 in
\epsfbox{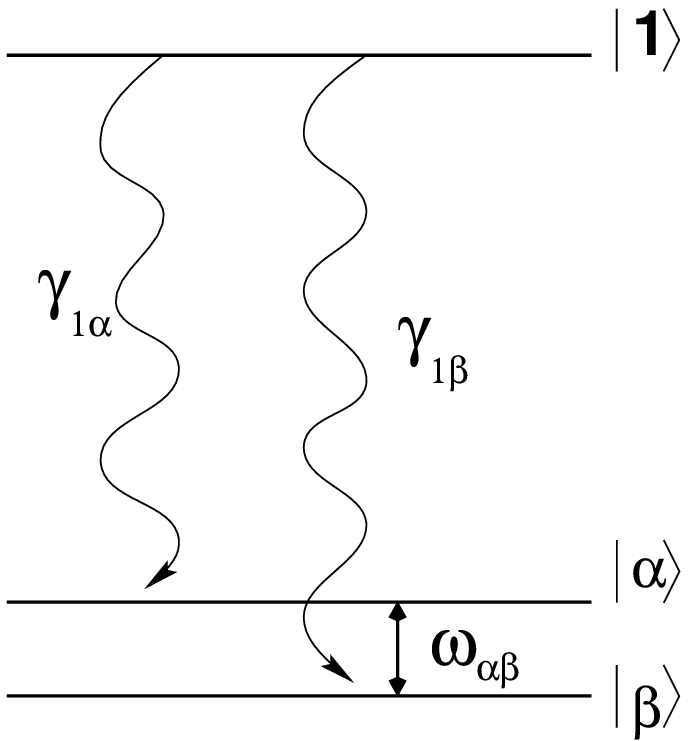}
\vspace*{0.5 cm}
\caption{Schematic diagram of a three level $\Lambda$ system.  The two
ground states
$|\alpha\rangle$ and $|\beta\rangle$ are coupled to the excited state $|1\rangle$ 
via vacuum field.} 
\label{one}
\end{figure}

Consider a $\Lambda$-system as shown in Fig. \ref{one}.
Let us take the zero of energy at the state $|\beta\rangle$ and let $\hbar \omega_{ij}$
denote the energy difference between the states $|i\rangle$ and $|j\rangle$. 
The Hamiltonian for this system interacting with the vacuum of radiation field is 
\begin{equation}
H = H_0 + H_{AV}.
\label{ham1}
\end{equation}
 where the unperturbed atom and vacuum field Hamiltonian $H_0$ will be
\begin{equation}
H_0 =  \hbar \omega_{1\beta} A_{11} + \hbar\omega_{\alpha\beta}
A_{\alpha\alpha} + \sum_{ks} \hbar\omega_k a^{\dagger}_{ks}a_{ks},
\label{ham2}
\end{equation}
and the interaction Hamiltonian $H_{AV}$ is 
\begin{equation} 
H_{AV} =  - \sum_{ks} \hbar\{(g_{ks} A_{1\alpha} +
f_{ks}A_{1\beta})a_{ks} + {\rm H.c}\}.
\label{ham3}
\end{equation}
Here the operator $A_{ij} = |i\rangle\langle j|$ is the atomic transition operator for
$i \neq j$ and population operator for $i = j$. The field annihilation  and creation
operators are $a_{ks}$  and $a^{\dagger}_{ks}$ respectively where the subscript denote the  
$k^{th}$ mode of the field with polarization along $\hat{\varepsilon}_{ks}$.
 The vacuum
coupling strengths are 
$g_{ks} = i(2\pi ck/\hbar L^3)^{1/2} \vec{d}_{1\alpha} \cdot \hat{\varepsilon}_{ks}e^{i\vec{k}\cdot \vec{r}}$ and 
$f_{ks} = i(2\pi ck/\hbar L^3)^{1/2} \vec{d}_{1\beta} \cdot \hat{\varepsilon}_{ks}e^{i\vec{k}
\cdot \vec{r}}$, where $\vec{d}_{1 i}$'s ($i = \alpha,\beta$) denote the dipole matrix 
elements.  We have dropped the anti-resonant terms from (\ref{ham3}).  
We work with density matrices and use the standard   
master equation technique here.  A calculation leads to the following  
master equation  for the reduced density matrix $\rho$ of the atomic system 
\begin{eqnarray}
\dot{\rho} &=& -i [\omega_{1\beta} A_{11} + \omega_{\alpha\beta}
A_{\alpha\alpha}, \rho] -\gamma_{1\alpha}(A_{11}\rho - 2A_{\alpha\alpha} \rho_{11}
+ \rho A_{11}) -\gamma_{1\beta}(A_{11}\rho \nonumber\\
& & - 2A_{\beta\beta}\rho_{11} + \rho A_{11})
+ 2\sqrt{\gamma_{1\alpha} \gamma_{1\beta}} \cos\theta_1
A_{\beta\alpha} \rho_{11}  + 2\sqrt{\gamma_{1\alpha}
\gamma_{1\beta}}
\cos\theta_1
A_{\alpha\beta} \rho_{11}.
\label{master}
\end{eqnarray}
Here $2\gamma_{1\alpha}= 4\omega_{1\alpha}^3|d_{1\alpha}|^2/3\hbar c^3$
and $2\gamma_{1\beta} = 4\omega_{1\beta}^3|d_{1\beta}|^2/3\hbar c^3$  denote
the spontaneous emission rates from state $|1\rangle$ to states $|\alpha\rangle$
and $|\beta\rangle$ respectively and
$\theta_1$ is the angle between the two transition dipole moments $\vec{d}_{1i}$ ($i =
\alpha, \beta$).
The last two terms in the above equation are the interference term due to coupling 
of the two atomic transition $|1\rangle \rightarrow |\alpha\rangle$, 
$|1\rangle \rightarrow |\beta\rangle$ to a common vacuum \cite{harris} 
of the electromagnetic
field.  The dipole matrix elements should be non-orthogonal for the above 
interference to occur.  The density matrix elements, $\rho_{ij}$ 
($\langle i|\rho|j\rangle$), in the Schr\"odinger picture obey equations  
\begin{eqnarray}
\dot{\rho}_{11} &=& -2\Gamma_1 \rho_{11},~~~~~~~~~~~~~~~
\dot{\rho}_{1\alpha} = -(\Gamma_1+i\omega_{1\alpha})
\rho_{1\alpha},\nonumber\\
\dot{\rho}_{\alpha\alpha} &=& 2\gamma_{1\alpha}\rho_{11},~~~~~~~~~~~~~~~~~
\dot{\rho}_{1\beta} =
-(\Gamma_1+i\omega_{1\beta})\rho_{1\beta},\nonumber\\
\dot{\rho}_{\beta\beta} &=& 2\gamma_{1\beta} \rho_{11},~~~~~~~~~~~~~~~~~
\dot{\rho}_{\alpha\beta} = -i\omega_{\alpha\beta} \rho_{\alpha\beta} + 
2\sqrt{\gamma_{1\alpha}\gamma_{1\beta}} \cos\theta_1 \rho_{11},
\label{elements}
\end{eqnarray}
where $\Gamma_1 = \gamma_{1\alpha} + \gamma_{1\beta}$.
Note that the equation for the ground state coherence $\rho_{\alpha\beta}$ is coupled
to the population of the excited state.  
Solving for the coherence $\rho_{\alpha\beta}$ with the initial condition $\rho (0) = A_{11}$ gives,
\begin{equation}
\rho_{\alpha\beta}(t) = \frac{2\sqrt{\gamma_{1\alpha}\gamma_{1\beta}}\cos\theta_1
(e^{-i\omega_{\alpha\beta} t} - e^{-2\Gamma_1 t})}
{(2\Gamma_1 - i\omega_{\alpha\beta})}~.
\label{coh0}
\end{equation}
As the equation reads, this coherence is non-zero as a result of interference term.
Even in the long time limit ($t \gg 1/\Gamma_1$ ) this coherence is finite 
and oscillates 
with a frequency $\omega_{\alpha\beta}$
\begin{equation}
\rho_{\alpha\beta}(t \rightarrow \infty) = \frac{2\sqrt{\gamma_{1\alpha}
\gamma_{1\beta}}\cos\theta_1e^{-i\omega_{\alpha\beta} t}}
{(2\Gamma_1 - i\omega_{\alpha\beta})}. 
\end{equation}
The magnitude of this coherence
is especially significant only if $\omega_{\alpha\beta} \le 2\Gamma_1$ \cite{olga} and if the dipole matrix elements are parallel.
Thus, as mentioned, even vacuum of electromagnetic field can give rise to coherence 
in systems with near degenerate levels.  We next address the questions: (a) what 
leads to the coherence (\ref{coh0}), and (b) how such a coherence can be measured.

In the long time limit the non-zero density matrix in (\ref{elements}) will be
\begin{eqnarray}
\rho_{\alpha\alpha} = \gamma_{1\alpha}/\Gamma_1,~~~ \rho_{\beta\beta} = \gamma_{1\beta}/\Gamma_1,~~~
\rho_{\alpha\beta} =  \sqrt{\rho_{\alpha\alpha}\rho_{\beta\beta}}B,~~~
{\rm where} ~~~~~
B = \frac{2\cos\theta_1}{(2-i\omega_{\alpha\beta}/\Gamma_1)}.
\label{elements2}
\end{eqnarray}
  The oscillation in $\rho_{\alpha\beta}$ 
has been removed by writing it in the interaction picture.  Thus the density matrix
$\rho$ be reduced to an
effective matrix $\tilde{\rho}$ where
\begin{equation} \tilde{\rho} = \left[
\begin{array}{ccc}
\rho_{\alpha\alpha} & \rho_{\alpha\beta}\\
\rho_{\beta\alpha} & \rho_{\beta\beta} 
\end{array}\right] \equiv \left[
\begin{array}{ccc}
\rho_{\alpha\alpha} & \sqrt{\rho_{\alpha\alpha}\rho_{\beta\beta}}B\\
\sqrt{\rho_{\alpha\alpha}\rho_{\beta\beta}}B^* & \rho_{\beta\beta} 
\end{array}\right].
\label{matrix}
\end{equation}
A measure of the purity of the state we calculate Tr($\tilde{\rho}^2)$ :
\begin{equation}
{\rm Tr}(\tilde{\rho}^2) = \frac{\gamma_{1\alpha}^2+\gamma_{1\beta}^2}
{\Gamma_1^2} + \frac{8\gamma_{1\alpha}\gamma_{1\beta}\cos^2\theta_1}
{(4\Gamma_1^2 + 
\omega^2_{\alpha\beta})}.
\label{trace1}
\end{equation}
For $\omega_{\alpha\beta} = 0$ and $\cos\theta_1 = 1$, $B = 1$, and we get
\begin{equation}
{\rm Tr}(\tilde{\rho}^2) = 1,
\label{trace2}
\end{equation}
which means that the atom would be in a pure state.  That is when the VIC
is maximum.  Generally, one would
find the system in a mixed state [Tr($\tilde{\rho}^2) < 1$] as $|B| \neq 1$.  The 
entropy of the final state depends on the parameter $B$.  

The case when the atom
is left in a pure state is especially interesting as we can introduce 
 the coupled ($|c\rangle$) and 
uncoupled ($|uc\rangle$) states given by
\begin{equation}
|c\rangle = \frac{|d_{1\alpha}||\alpha\rangle + |d_{1\beta}||\beta\rangle}{|d|},
~~~~~~|uc\rangle = \frac{|d_{1\beta}||\alpha\rangle - |d_{1\alpha}||\beta
\rangle}{|d|},
\end{equation}
where $|d| = \sqrt{|d_{1\alpha}|^2 + |d_{1\beta}|^2}$.
 The Hamiltonian (\ref{ham1}) can be written as  
\begin{equation}
H = \hbar\omega_{1\beta} |1\rangle\langle 1| + 
 \sum_{ks}
\omega_k a^{\dagger}_{ks}a_{ks} - \sum_{ks} (g^{\prime}_{ks}|1\rangle\langle c| a_{ks} + 
{\rm H.c}),
\label{ham4}
\end{equation}
where $g^{\prime}_{ks} = i(2\pi ck/\hbar L^3)^{1/2}|d|\hat{d}\cdot\hat
{\varepsilon}_{ks} e^{i\vec{k}\cdot\vec{r}}$ is the vacuum coupling between 
state $|1\rangle$ and $|c\rangle$ and $\hat{d}$ is the unit vector parallel
to both $\vec{d}_{1\alpha}$ and $\vec{d}_{1\beta}$.  Note that state 
$|uc\rangle$ is not directly coupled to state $|1\rangle$.  Thus $|uc\rangle$
never gets populated if $\rho(0) = A_{11}$.  The spontaneous
emission from state $|1\rangle$ occurs to the coherent superposition 
state $|c\rangle$ and {\em not just} the individual states $|\alpha\rangle$ 
and $|\beta\rangle$.  Clearly under these conditions the final state will be $|c\rangle$
which agrees with the result (\ref{elements2}) for $\omega_{\alpha\beta} = 0$, 
$\theta_1 = 0$.  

For $\omega_{\alpha\beta} \neq 0$, the proper basis corresponds to the two
eigenstates $|\psi_{\pm}\rangle$ of (\ref{matrix}) and the steady state will be
an {\em incoherent} mixture of $|\psi_+\rangle$ and $|\psi_-\rangle$.

\section{Emission spectrum}
We now come to the question as to how can one probe the existence of VIC in a 
$\Lambda$-system.  Thus we naturally think of the spectrum of spontaneous emission.
In a V-system the spontaneous emission is significantly affected
by the presence of VIC
\cite{{gsa1},{stroud},{agassi},{plenio},{zhu},{xia},{gsa2},{swain}}.  But for a $\Lambda$-system, as we show, the  emission
spectrum is independent of VIC.  The emission spectrum corresponds to the normally ordered
two time correlation function of electric field amplitudes \cite{gsa1}.  The radiated fields at spacetime points
$\vec{r}_l,t_l$ ($l = 1,2$) will have a correlation given by  
\begin{equation}
\langle E^{(-)}(\vec{r}_1,t_1)\cdot E^{(+)}(\vec{r}_2,t_2)\rangle = (r_1r_2)^{-1}
\sum_{i,j = \alpha, \beta} M_{ij}
\langle A_{1i}(t_1) A_{j1}(t_2)\rangle,~~~~~~~~t_1 > t_2,  
\label{field1}
\end{equation}
\begin{eqnarray*}
{\rm where}~~~~~~~
M_{ij} = (\frac{\omega_{1i}\omega_{1j}}{c^2})^2
[\hat{r}_1\times(\hat{r}_1\times\vec{d}^*_{1i})]\cdot
[\hat{r}_2\times(\hat{r}_2\times\vec{d}_{1j})],
\end{eqnarray*}
and $r_l$ is much greater than the size of the source.
Using quantum regression theorem and equations (\ref{elements}), it can be shown that
the two time 
atomic correlation functions are given by  
\begin{equation}
\langle A_{1i}(t_1) A_{i1}(t_2) \rangle = \exp{[(i\omega_{1i} -\Gamma_1) 
(t_1 - t_2)]} \exp{(-2\Gamma_1t_2)},~~~~~~t_1 > t_2,
\label{corr1}
\end{equation}
\begin{equation}
{\rm and}~~~~~ \langle A_{1i}(t_1) A_{j1}(t_2) \rangle = 0~~~~~ {\rm for}~~~~~i \neq j. 
\end{equation}
Using (\ref{corr1}) in (\ref{field1}) we get the correlation function of the radiated
field  
\begin{equation}
\langle E^{(-)}(\vec{r}_1,t_1)\cdot E^{(+)}(\vec{r}_2,t_2)\rangle = (r_1r_2)^{-1}
\sum_{i = \alpha, \beta} M_{ii}
 \exp{[(i\omega_{1i} -\Gamma_1)
(t_1 - t_2)]} \exp{(-2\Gamma_1t_2)},~~~~t_1 > t_2.
\end{equation}
This correlation function is 
the sum of {\em incoherent} emissions along the two transitions, $|1\rangle \rightarrow 
|\alpha\rangle$, $|1\rangle \rightarrow |\beta\rangle$.  Thus we conclude that the
spontaneous emission spectrum 
in a $\Lambda$-system is {\em not affected} by VIC.  Therefore one has to consider other
types of probes to study VIC in such a system.

\section{Modulated absorption as a probe of VIC}
\begin{figure}[h]
\hspace*{0.2 cm}
%\vspace*{}
%\epsfxsize  4.5 in
%\epsfysize 2.5 in
\epsfbox{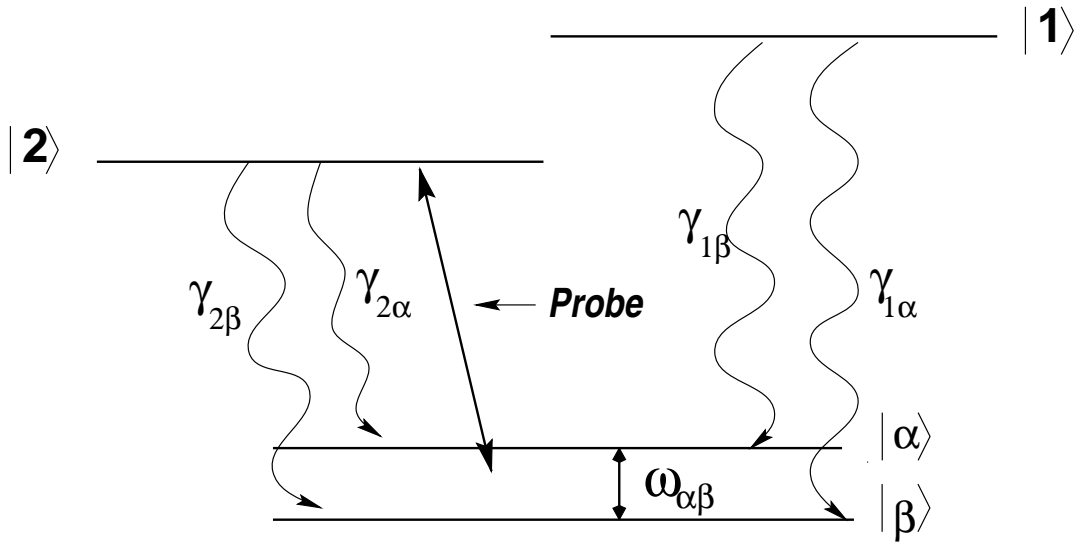}
\vspace*{0.5 cm}
\caption{Schematic diagram of a four level model proposed for monitoring vacuum induced
coherence.  The
coherence created after spontaneous emission from $|1\rangle$ can be observed in the probe
absorption.}
\label{two}
\end{figure}

The above result is not surprising because the coherence is created after the 
spontaneous emission has occurred.  An alternative approach to monitor VIC will be to
study the absorption of a probe field tuned close to some other transition in the 
system.  In this paper we show that 
a unique feature in probe absorption appears
due to the presence of VIC.  The model scheme is as 
shown in Fig. \ref{two}.  Here the spontaneous emission from state $|1\rangle$
creates VIC between the two near-degenerate ground levels $|\alpha\rangle$
 and $|\beta\rangle$.  We now consider another excited state $|2\rangle$, well separated 
from $|1\rangle$.  A weak coherent field is tuned between
state $|2\rangle$ and the two ground states  
to monitor VIC.  
The Hamiltonian in the dipole approximation will be
\begin{equation}
{\cal H} = \hbar\omega_{\alpha\beta} A_{\alpha\alpha} + \hbar\omega_{1\beta} A_{11}
+ \hbar\omega_{2\beta} A_{22} - \{(\vec{d}_{2\beta} A_{2\beta} + 
\vec{d}_{2 \alpha} A_{2\alpha})\cdot 
\vec{E}_2 e^{-i\omega_2 t} + {\rm H.c}\},
\label{ham}
\end{equation}
where the  counter rotating terms in the probe field have been dropped.
The probe field is treated classically here and 
has a frequency $\omega_2$
and a complex amplitude $\vec{E}_2$.  We use the master equation method to derive
equations for the reduced density matrix of the atomic system.  We give the
result of such a calculation,
\begin{mathletters}
\begin{eqnarray}
\dot{\rho}_{11} &=& -2\Gamma_1 \rho_{11},\\
\dot{\rho}_{22} &=& -2\Gamma_2 \rho_{22} + i(G\rho_{\alpha 2} 
+ F\rho_{\beta 2})e^{-i\omega_2 t} -i(G^*\rho_{2 \alpha} + F^*\rho_{2 \beta})
e^{i\omega_2 t},
\label{b}\\
\dot{\rho}_{\alpha \alpha} &=& 2\gamma_{1\alpha} \rho_{11} + 2\gamma_{2\alpha} \rho_{22} 
-iG e^{-i\omega_2 t}\rho_{\alpha 2} + iG^* e^{i\omega_2 t}\rho_{2 \alpha},
\label{c}\\
\dot{\rho}_{\alpha 2} &=& -(\Gamma_2 - i\omega_{2\alpha})
\rho_{\alpha 2} - iF^*e^{i\omega_2 t}\rho_{\alpha \beta} + iG^*e^{i\omega_2 t}(\rho_{22} - 
\rho_{\alpha \alpha}),\label{d}\\
\dot{\rho}_{\beta 2} &=& -(\Gamma_2 - i\omega_{2\beta})
\rho_{\beta 2} - iG^* e^{i\omega_2 t} \rho_{\beta\alpha} + iF^* e^{i\omega_2 t} (2\rho_{22} 
+ \rho_{11} + \rho_{\alpha\alpha}- 1),\label{e}\\
\dot{\rho}_{\alpha 1} &=& -(\Gamma_1 - i\omega_{1\alpha})\rho_{\alpha
1} 
+ iG^* e^{i\omega_2 t}\rho_{21},\\
\dot{\rho}_{\beta 1} &=& -(\Gamma_1 - i\omega_{1\beta})
\rho_{\beta 1} + iF^* e^{i\omega_2 t}\rho_{21},\\
\dot{\rho}_{21} &=& -(\Gamma_1 + \Gamma_2
-i(\omega_{1\beta}-\omega_{2\beta}))\rho_{21}
+i(G\rho_{\alpha 1} + F\rho_{\beta 1}) e^{-i\omega_2 t},\\
\dot{\rho}_{\alpha \beta} &=& \eta_1 \rho_{11} + \eta_2 \rho_{22} -i\omega_{\alpha\beta} 
\rho_{\alpha \beta} - iF e^{-i\omega_2 t} \rho_{\alpha 2} + iG^* e^{i\omega_2 t}\rho_{2 \beta},
\label{density}
\end{eqnarray} 
\end{mathletters}
where we have used the trace condition $\sum_i \rho_{ii} = 1$ for (\ref{e}).
Here 
\begin{equation}
2\gamma_{2\alpha} = \frac{4\omega_{2\alpha}^3 |d_{1\alpha}|^2}{3\hbar c^3}~~~
{\rm and}~~~ 
2\gamma_{2\beta} = \frac{4\omega_{2\beta}^3 |d_{1\beta}|^2}{3\hbar c^3} 
\end{equation}
define the spontaneous emission rates from $|2\rangle$ to states $|\alpha\rangle$
and $|\beta\rangle$ respectively and we write $\Gamma_2 =
\gamma_{2\alpha} + \gamma_{2\beta}$.  The Rabi frequencies 
\begin{equation}
2G = 2\vec{E}_2 \cdot \vec{d}_{2 \alpha}/\hbar, ~~~~2F = 2\vec{E}_2 \cdot 
\vec{d}_{2 \beta}/\hbar
\end{equation}
are for the probe field acting on transitions $|1\rangle \leftrightarrow |\alpha\rangle$
and $|1\rangle \leftrightarrow |\beta\rangle$ respectively.  Further
we can write $G = |G|e^{-i\phi_1}$ and $F = |F|e^{-i\phi_2}$, where the phase
$\phi = \phi_1 - \phi_2$ gives the relative phase
between the complex dipole matrix elements $\vec{d}_{1\alpha}$ and $\vec{d}_{1\beta}$.  
The VIC parameters are 
\begin{equation}
\eta_1 = 2\sqrt{\gamma_{1\alpha} \gamma_{1\beta}}\cos \theta_1, ~~~~~~
\eta_2 = 2\sqrt{\gamma_{2\alpha} \gamma_{2\beta}}\cos \theta_2.   
\end{equation}
We thus include vacuum induced coherence on all possible transitions.

In order to study probe absorption we solve Eqs. (19) perturbatively.  We need to know
$\rho_{22}(t)$ to second order in the probe field, assuming that the atom was prepared
in the state $|1\rangle$ at $t = 0$. 
Using (\ref{b}) we get
\begin{equation}
\rho_{22}^{(2)} (t) = i\int_0^t d\tau e^{-i\omega_2 \tau} [|G|e^{-i\phi_1} 
\rho_{\alpha 2}^{(1)}(\tau) + |F|e^{-i\phi_2} \rho_{\beta 2}^{(1)} (\tau)] 
e^{-2\Gamma_2(t-\tau)} + {\rm c.c.}~~.
\label{second1}
\end{equation}
The first order contribution is obtained, for example, by integrating (\ref{d})
\begin{equation}
\rho_{\alpha 2}^{(1)} (t) = -i\int_0^t d\tau e^{i\omega_2 \tau} \{|F|e^{i\phi_2} 
\rho_{\alpha \beta}^{(0)} (\tau) - |G|e^{i\phi_1} [\rho_{22}^{(0)}(\tau) -
\rho_{\alpha\alpha}^{(0)}(\tau)]\} 
e^{-(\Gamma_2 -i\omega_{2\alpha})(t-\tau)},
\label{first}
\end{equation}
 It can be easily show
that $\rho_{22}^{(0)}(t) = 0$ and the other zeroth order terms are known from Sec. I.  
The VIC contribution arises from non-zero $\rho_{\alpha\beta}^{(0)}(t)$ in (\ref{first}).
Similarly integrating for $\rho_{\beta 2}^{(1)}(t)$ and combining with Eqs. (\ref{first})
(\ref{second1}), and on simplification we find our key result
\begin{eqnarray}
\rho_{22}^{(2)}(t \gg \Gamma_1^{-1}, \Gamma_2^{-1}) &\equiv& \frac{\eta_1 |F||G| e^{-i(\omega_{\alpha\beta}
 t + \phi)}}{(2\Gamma_2-i\omega_{\alpha\beta})(2\Gamma_1-i\omega_{\alpha\beta})(\Gamma_2 
-i(\Delta_2+\omega_{\alpha\beta}/2))} \nonumber\\
& &+ \frac{\eta_1 |F||G| e^{i(\omega_{\alpha\beta}
 t + \phi)}}{(2\Gamma_2+i\omega_{\alpha\beta})(2\Gamma_1+i\omega_{\alpha\beta})(\Gamma_2
-i(\Delta_2-\omega_{\alpha\beta}/2))} \nonumber\\
& &+  \frac{|G|^2}{4\Gamma_2[\Gamma_2 - 
i(\Delta_2 - \omega_{\alpha\beta}/2)]} + 
\frac{|F|^2}{4\Gamma_2[\Gamma_2 - i(\Delta_2 + \omega_{\alpha\beta}/2)]} + {\rm c.c}.
\label{second2}
\end{eqnarray}
When $\eta_1 \rightarrow 0$

\begin{equation}
 \rho_{22}^{(2)}(t \gg \Gamma_1^{-1}, \Gamma_2^{-1}) \equiv \frac{|G|^2}{4\Gamma_2[\Gamma_2 - i(\Delta_2 -
 \omega_{\alpha\beta}/2)]} + 
\frac{|F|^2}{4\Gamma_2[\Gamma_2 - i(\Delta_2 + \omega_{\alpha\beta}/2)]} + {\rm c.c},
\label{second3}
\end{equation}
where the last result is the expected result which is the sum of the individual
absorptions corresponding to the transitions $|\alpha\rangle \rightarrow |2\rangle$,
$|\beta\rangle \rightarrow |2\rangle$.  The parameter $\Delta_2 = \omega_{1\beta} -
\omega_{\alpha\beta}/2 - \omega_2$ is the probe detuning defined with respect to the
center of level $|\alpha\rangle$ and $|\beta\rangle$.  The modulated term in probe 
absorption (\ref{second2}) is the
result of VIC.  {\em This modulation is the signature of the VIC produced by the
two paths of spontaneous emission} $|1\rangle \rightarrow |\alpha\rangle$, 
$|1\rangle \rightarrow |\beta\rangle$.  Note the interesting phase dependence
that arises in the probe absorption due to non-zero $\eta_1$.  This phase dependence
is another outcome of the presence of VIC in a system.   Since the probe is 
treated to second
order in its amplitude, the result is independent of the coherence parameter
$\eta_2$ for the transition $|2\rangle\rightarrow |\alpha\rangle$, $|2\rangle
 \rightarrow |\beta\rangle$.  Needless to say that the Eqs. (19) can 
be integrated numerically to obtain the probe absorption for arbitrary times.  For
this purpose it is useful to remove the optical frequencies by making
 the transformations $\tilde{\rho}_{1i} \equiv \rho_{1i} e^{i\omega_{1i}t}$,
$\tilde{\rho}_{2i} \equiv \rho_{2i} e^{i\omega_{2}t}$ ($i = \alpha, \beta$) and 
$\tilde{\rho}_{12} \equiv \rho_{12}e^{i(\omega_2 - \omega_{1\beta})t}$  etc. 
We solve these 
using fifth-order Runge-Kutta-Verner method with the 
initial condition that $\rho_{11}(0) = 1$.  We take the probe Rabi frequencies 
$F,G$ much smaller than $\Gamma$'s.  The numerical results for excited state population
$\rho_{22}(t)$ as a function of time for both the cases when $\eta_1$ is zero and non-zero
are plotted in Fig. 3.  Figure 3 shows the significant difference that arises due to 
the presence or absence of VIC.  The oscillation in the probe absorption is the reflection
of oscillation of in the coherence $\rho_{\alpha\beta}$ (see (\ref{coh0}))  and 
this confirms the analytical result (\ref{second2}).  The numerical result shows
a very slow decay of the envelop of the oscillations.  This arises from terms
which are of higher order in probe strength. 

Finally we discuss the changes in absorption spectrum that can arise due to VIC.
The modulated component of the population (\ref{second2}) can be written as
\begin{eqnarray}
\rho_{22}^{(2)} &\equiv& \frac{2\eta_1 |F||G|}{D}[\{2\Gamma_1 \Gamma_2^2 + 2\Gamma_1
(\Delta_2^2-
\omega_{\alpha\beta}^2/4) - \Gamma_2 \omega_{\alpha\beta}^2\}\cos(\omega_{\alpha\beta}t
+\phi) \nonumber\\
& &+ \omega_{\alpha\beta}\{2\Gamma_1\Gamma_2 + \Gamma_2^2 + \Delta_2^2-
\omega_{\alpha\beta}^2/4\}\sin(\omega_{\alpha\beta}t + \phi)],
\label{second4}
\end{eqnarray}
where
\begin{eqnarray*}
D = (4\Gamma_1^2+\omega_{\alpha\beta}^2)[\Gamma_2^2+(\Delta_2+\omega_{\alpha\beta}/2)^2]
[\Gamma_2^2+(\Delta_2-\omega_{\alpha\beta}/2)^2].
\end{eqnarray*}
Since it is possible to separate the sine and cosine terms by a phase sensitive
detection we plot these in Fig. 4
as a function of probe detuning.   These two components of the absorption spectrum
behave quite differently.

\section{conclusions}
The important criteria for the existence of the vacuum induced coherence between the 
close lying levels is the nonorthogonality of the dipole matrix elements.  In practice
the nonorthogonality can be achieved by mixing of the energy levels.  The mixing
can occur either due to internal fields or due to externally applied fields.  For
example in the experiment of Xia {\it et at.} \cite{xia}, spin-orbit interaction
gives rise to mixing.  The VIC has also been studied when the level mixing is 
produced by using
electromagnetic fields \cite{anil1}, dc fields \cite{berman}, and rf fields
\cite{lenstra}.  Special configurations involving cavities can also be utilized to
study VIC \cite{anil2}.

In conclusion we have found that the VIC in a $\Lambda$-system is more difficult
to monitor as it does not show up in the fluorescence spectrum.  We have however
demonstrated that the absorption spectrum carries the information on VIC and that the VIC
produces a modulated component in the absorption spectrum.

\newpage
\begin{center}
FIGURES
\end{center}

\newpage
\begin{figure}
\hspace*{0.7 cm}
%\vspace*{}
\epsfxsize 5in
\epsfysize 6in
\epsfbox{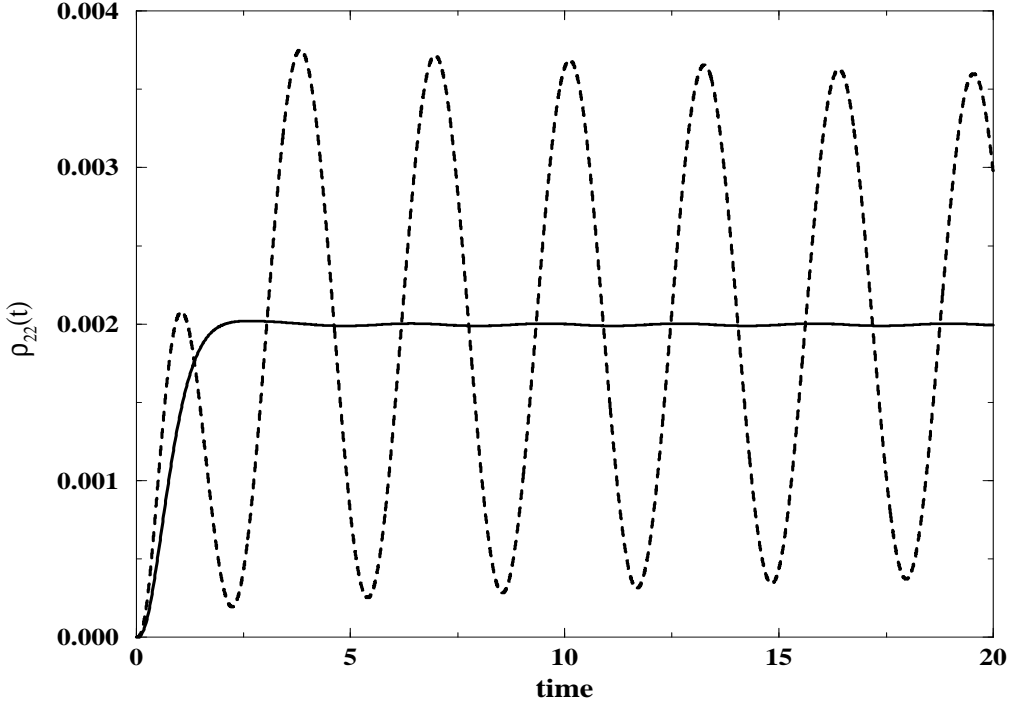}
\vspace*{-1 cm}
\caption{The excited state population $\rho_{22}$ as a function of
scaled time $\gamma t$.  Here the probe field is tuned
to the center of states $|\alpha\rangle$ and $|\beta\rangle$, and we take
$\gamma_{1\alpha} = \gamma_{1\beta} = \gamma_{2\alpha} = \gamma_{2\beta} = \gamma$.  
The parameters are chosen as $F/\gamma = G/\gamma = 0.1$ and $\omega_{\alpha\beta}/\gamma = 2$,
and phase $\phi = 0$.
The dashed oscillating curve is in the presence of VIC and the solid
curve in the absence of VIC.  This observed modulation is consistent with the analytical
result (\ref{second2}).  Very weak oscillation 
appears in the solid line because
the probe is not exactly tuned to the two transitions.}
\label{three}
\end{figure}

\newpage
\begin{figure}
\hspace*{0.7 cm}
%\vspace*{}
\epsfxsize 5in
\epsfysize 6.3in
\epsfbox{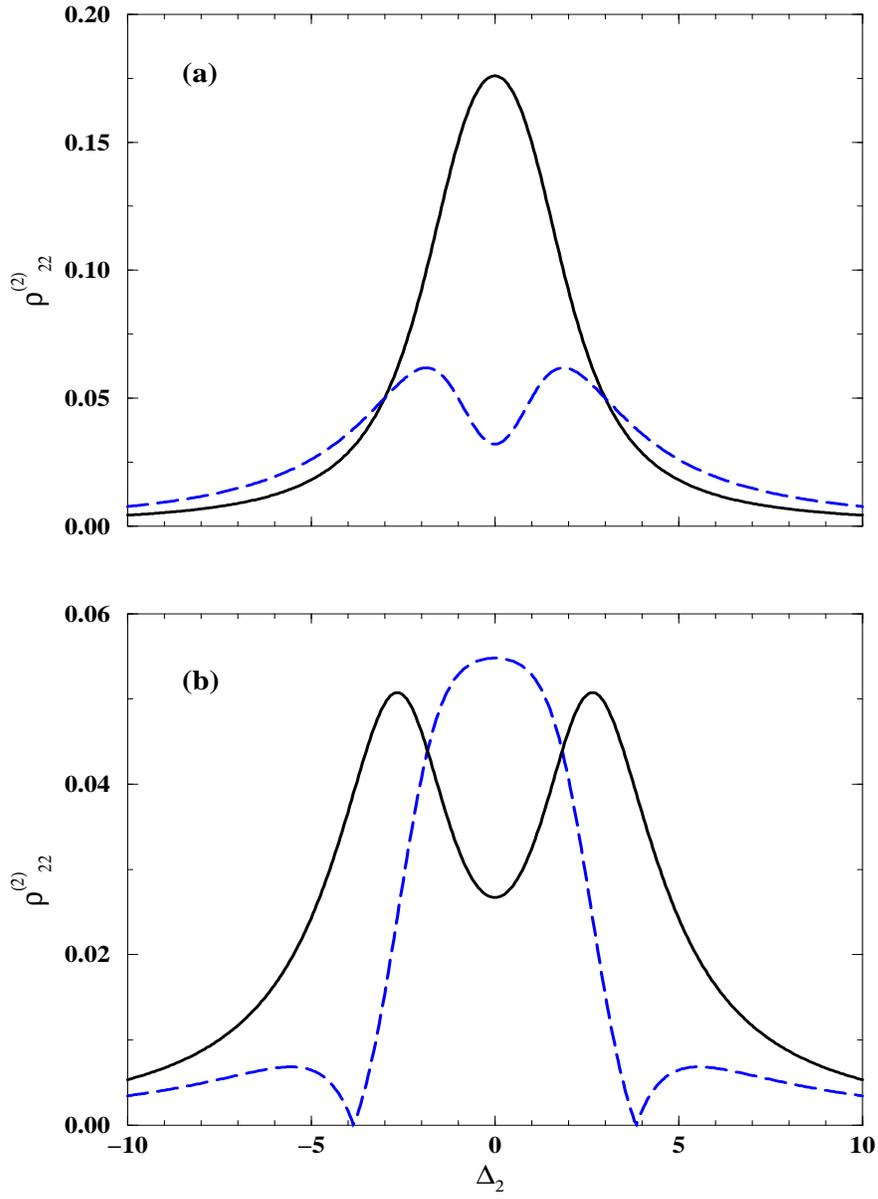}
%\vspace*{-1 cm}
\caption{The cosine (dashed) and sine (solid) components of the excited state
population $\rho_{22}^{(2)} \times 10^2$ as a function of probe detuning.  Plot
(a) is for $\omega_{\alpha\beta} = 2\gamma$ and (b) is for $\omega_{\alpha\beta} = 5\gamma$.    
The parameters are as in Fig. (\ref{three}).}
\label{four}
\end{figure}

\end{document}